\documentclass[preprint,aps]{revtex4}

\usepackage{graphicx}
\usepackage{dcolumn}
\usepackage{bm}
\usepackage{psfig}

\begin{document}


\title{Ultralong coherence times in the purely electronic zero-phonon line emission of single molecules}

\author{Alper Kiraz, Moritz Ehrl, Christoph Br\"{a}uchle, and Andreas Zumbusch}

\affiliation{Department Chemie, Ludwig-Maximilians Universit{\"a}t M{\"u}nchen,
Butenandtstr. 11, D-81377 M{\"u}nchen, Germany}

\date{\today}

\begin{abstract}

We report the observation of ultralong coherence times in the purely electronic zero-phonon line emission 
of single terrylenediimide molecules at 1.4~K. Vibronic excitation
and spectrally resolved detection with a scanning Fabry-Perot spectrum analyzer were used
to measure a linewidth of 65~MHz. This is within a factor of 1.6 of the transform limit. It therefore indicates that
single molecule emission may be suited for applications in linear optics quantum computation. Additionally
it is shown that high resolution spectra taken with the spectrum analyzer allow for the investigation 
of fast spectral dynamics in the emission of a single molecule.

\end{abstract}


\maketitle

Schemes employing linear optics offer a promising approach for the realization of efficient
quantum computation~\cite{knill2001}. True single photon sources generating photons on demand
are indispensable for this purpose. Until recently, such ideal light sources have been substituted 
with parametric down conversion in nonlinear crystals generating twin photons. Even though high levels 
of indistinguishability, and thus strong Hong-Ou-Mandel correlations~\cite{hong1987}, can be achieved 
with this method, experiments using this source are bound to be performed 
at very low intensities. This however impedes the implementation of large scale protocols~\cite{sanaka2004}. 

Single photon sources based on two-level emitters constitute an attractive alternative to twin 
photons. Single molecules~\cite{demartini1996,brunel1999,lounis2000}, single quantum dots~\cite{michler2000,santori2001}, 
and single nitrogen-vacancies in diamond~\cite{beveratos2002a} have successfully been used as such sources. 
They can provide single photons at high emission rates with negligible two or more photon generation probabilities. A 
crucial remaining requirement for their application to linear optics quantum information protocols is
photon indistinguishability.

Indistinguishability of emitted photons from incoherently excited two-level emitters is directly related to their 
coherence length, provided that the excited level is populated via a fast relaxation process~\cite{kiraz2004}.
In the case of transform limited emission, a high level of indistinguishability 
is then granted. The ultimate transform limit is given by $2T_1/T_2=1$, where $T_1$ and $T_2$ represent 
the population and coherence decay times, respectively. This limit has recently been approached within a 
minimum factor of $2T_1/T_2\sim 1.5$ for single InAs quantum dots~\cite{bayer2002,kammerer2002,santori2002}. 
These results enabled direct measurements of 
indistinguishability~\cite{santori2002}, violation of Bell's inequality~\cite{fattal2004a}, and quantum 
teleportation~\cite{fattal2004b}. In contrast, the linewidth measurement of a single nitrogen-vacancy in diamond revealed 
a coherence time which was much smaller than the transform limit~\cite{jelezko2003}.

For more than a decade, single molecules in organic host materials have been detected at cryogenic 
temperatures~\cite{tamarat2000,moerner2002}. With one exception~\cite{nonn2001}, these experiments 
rely on the narrow bandwidth excitation of the molecule's purely electronic zero-phonon line (ZPL). The 
detected Stokes shifted fluorescence is composed of the contributions from several vibrational states, 
but does not contain the ZPL. In these experiments, transform limited absorption of the ZPL is routinely observed by scanning 
the excitation wavelength. To date however, the ZPL emission as a potential source of indistinguishable single 
photons has not been detected and analyzed.

Here, we demonstrate that nearly transform limited ZPL emission from single molecules can be obtained by using vibronic excitation. 
For this purpose, we excite vibronic transitions of individual molecules (Figure~1(a))~\cite{kiraz2003}. 
The fluorescence signal is now composed of the ZPL and the vibrational bands with
intensities determined by the respective Franck-Condon factors. From the integral fluorescence, 
the ZPL of single molecules can be selected with a spectrally narrow interference 
filter. The linewidth of the ZPL is measured by using a scanning Fabry-Perot 
spectrum analyzer (SA). Our excitation scheme is also less susceptible to spectral jumps of the molecules. 
The absorption bands have a spectral width of several cm$^{-1}$ which is determined by the lifetime of the excited 
vibrational state (1-10~ps) and its intense phonon sideband. For this reason, small spectral jumps of the 
molecular absorption do not lead to a loss of the excitation.

The experiments were performed using a home-built confocal microscope keeping the sample in a superfluid Helium bath 
at 1.4~K (Fig.~1(b)). The beam from a single mode dye laser (Coherent, 899-29) is reflected off a dichroic mirror, passed 
through a $\lambda/2$ plate, a galvo optic scanner and a telecentric system, before being focused onto the sample with 
an aspherical lens (NA=0.55), which is also used for collecting the fluorescence. After transmision through the $\lambda/2$ plate
and the dichroic mirror, the collected fluorescence is focused onto a pinhole (200~$\mu$m diameter) and dispersed with 
a 46~cm monochromator (Jobin-Yvon, HR460) before detection with a charge coupled device detector (CCD, Acton Research, 128HB).
Alternatively, the signal can be sent to the Fabry-Perot SA (Coherent, 240) before detection with an avalanche photodiode 
(APD, Perkin-Elmer, SPCM-AQR-16). A spectral resolution of 30~GHz is obtained in the spectra taken with the 
monochromator and the CCD. By contrast, the SA has a free spectral range of 1.5~GHz. Its finesse and peak transmission are 
measured to be $\sim$100 (spectral resolution $\sim$15~MHz), and $\sim$5\%, respectively. A broadband filter 
(Chroma, HQ 655/60), a color filter (Schott, RG630), and a narrowband interference filter (Omega, 677 SC2,
FWHM=2~nm, angle tunable between 650-677~nm) are used in spectrally selecting the ZPL emission. A gated counter 
(Stanford Instruments, SR400) is used for both external triggering of the SA and accumulation of the APD counts.

We examined terrylenediimide (TDI) molecules (Fig.~1(c)), highly diluted in the \mbox{Shpol'skii} matrix hexadecane. In this 
crystalline host, TDI molecules exhibit high spectral stabilities allowing the observation of lifetime limited ZPL absorption 
bands at 1.4~K~\cite{mais1999}. The sample solution was prepared by predissolving TDI molecules in a small amount of CH$_2$Cl$_2$. 
In order to remove the solvent CH$_2$Cl$_2$ and oxygen from the solution, several evaporation (10$^{-5}$~mbar) 
and freezing (77~K) cycles were performed. The sample was subsequently heated and quickly loaded to the cryostat.

In order to select a single molecule during an experiment, we first scan a 100~$\mu$m x 100~$\mu$m area at a fixed
excitation wavelength. We then choose the optimum excitation polarization at a sample location 
and scan the laser wavelength over the $\sim$10~nm width of 
the inhomogeneous band. By this, we can optimize the excitation conditions to detect the resolution limited ZPL
emission spectra of single TDI molecules on the CCD, as shown in the inset of Fig.~2(a).

High resolution spectra are obtained by directing this ZPL emission to the SA. A slow scan of the SA reveals spectra of 
the type depicted in Fig.~2(a). This scan consists of 600 bins, each of which corresponds to different mirror 
separations of the Fabry-Perot SA. At each bin, the APD counts are integrated for 500~ms. The excitation intensity
is 35~$\mu$W, which is much lower than the saturation power. For this reason, no power broadening is expected.
The Lorentzian modulations seen in the figure are separated by 1.5 GHz, which equals the free-spectral range
of the SA. A fit to the data yields a $65\pm10$~MHz linewidth for the ZPL emission. This corresponds to a 
coherence decay time of $4.9\pm0.9$~ns. By comparison, conventional fluorescence excitation spectroscopy of single TDI
molecules in hexadecane yield a homogeneous linewidth of $\sim$40 MHz~\cite{mais1999}. Therefore, our measured emission 
linewidth is within a factor of 1.6 of the transform limit. Such deviations from the transform limited value can be due 
to dynamic host-guest interactions occurring during the acquisition time at 1.4~K. These dynamics can be
increased by local heating induced by the excitation laser. Note, however, that local environments of individual molecules 
may also lead to homogeneous linewidth values differing from the ensemble value of $\sim$40~MHz.

Two experimental observations give conclusive evidence that the emission shown in Fig.~2(a) stems from a single molecule. First of
all, at the end of this experiment, we increased the excitation power to 10~mW and were able to thermally activate local 
degrees of freedom by laser induced heating. This resulted in blinking and a digital spectral jump in the fluorescence 
spectra (Fig.~2(b)). Secondly, given the large inhomogeneous broadening of the TDI band in hexadecane, it is highly 
improbable that several molecules would appear at the same spectral position in the SA.

We have observed Lorentzian ZPLs with narrow spectra for 8 other molecules. Apart from three molecules, 
which were spectrally stable over a time of 10 minutes, the other
molecules exhibited spectral dynamics even at moderate excitation intensities.
This is shown in Fig.~3, where we plot four fast consecutive scans taken with the ZPL fluorescence from such a molecule, 
at an excitation intensity of 200~$\mu$W. At each bin, an integration time of 25~ms was used. From the spectra it is
evident that the molecule preferentially occupies several spectral positions. Among those, at least two can be identified 
as indicated by the straight and dotted lines in the figure. It is however important to note that even molecules which
exhibit such spectral dynamics can be suitable sources of indistinguishable single photons. The only requirement which
needs to be fulfilled then is that they are spectrally stable on the time scale of the spontaneous emission 
lifetime. In our case, the spectral jumps take place on a ms time scale whereas the fluorescence lifetime is
10$^6$ times shorter.

In conclusion, we have observed nearly transform limited ZPL emission from a single TDI molecule
in hexadecane with a coherence decay time of 4.9~ns. We used a newly developed vibronic excitation at cryogenic 
temperatures in connection with a high resolution Fabry-Perot SA to measure these
linewidhts in emission. The results show that single molecules can serve as sources for indistinguishable
single photons. Our approach offers new possibilities for performing linear optics quantum information 
experiments like those reported in~\cite{santori2002,fattal2004a,fattal2004b} using larger coherence times.
It can also be employed to follow spectral dynamics of single molecules in emission.

The authors thank K. M\"{u}llen for a gift of TDI and T.H.P. Brotosudarmo for help with the sample
preparation. This work was supported by the Deutsche Forschungsgemeinschaft, SFB533, and the
Alexander von Humboldt Foundation (A.K.).

\newpage

\begin{thebibliography}{21}
\expandafter\ifx\csname natexlab\endcsname\relax\def\natexlab#1{#1}\fi
\expandafter\ifx\csname bibnamefont\endcsname\relax
  \def\bibnamefont#1{#1}\fi
\expandafter\ifx\csname bibfnamefont\endcsname\relax
  \def\bibfnamefont#1{#1}\fi
\expandafter\ifx\csname citenamefont\endcsname\relax
  \def\citenamefont#1{#1}\fi
\expandafter\ifx\csname url\endcsname\relax
  \def\url#1{\texttt{#1}}\fi
\expandafter\ifx\csname urlprefix\endcsname\relax\def\urlprefix{URL }\fi
\providecommand{\bibinfo}[2]{#2}
\providecommand{\eprint}[2][]{\url{#2}}

\bibitem[{\citenamefont{Knill et~al.}(2001)\citenamefont{Knill, Laflamme, and
  Milburn}}]{knill2001}
\bibinfo{author}{\bibfnamefont{E.}~\bibnamefont{Knill}},
  \bibinfo{author}{\bibfnamefont{R.}~\bibnamefont{Laflamme}}, \bibnamefont{and}
  \bibinfo{author}{\bibfnamefont{G.~J.} \bibnamefont{Milburn}},
  \bibinfo{journal}{Nature (London)} \textbf{\bibinfo{volume}{409}},
  \bibinfo{pages}{46} (\bibinfo{year}{2001}).

\bibitem[{\citenamefont{Hong et~al.}(1987)\citenamefont{Hong, Ou, and
  Mandel}}]{hong1987}
\bibinfo{author}{\bibfnamefont{C.~K.} \bibnamefont{Hong}},
  \bibinfo{author}{\bibfnamefont{Z.~Y.} \bibnamefont{Ou}}, \bibnamefont{and}
  \bibinfo{author}{\bibfnamefont{L.}~\bibnamefont{Mandel}},
  \bibinfo{journal}{Phys. Rev. Lett.} \textbf{\bibinfo{volume}{59}},
  \bibinfo{pages}{2044} (\bibinfo{year}{1987}).

\bibitem[{\citenamefont{Sanaka et~al.}(2004)\citenamefont{Sanaka, Jennewein,
  Pan, Resch, and Zeilinger}}]{sanaka2004}
\bibinfo{author}{\bibfnamefont{K.}~\bibnamefont{Sanaka}},
  \bibinfo{author}{\bibfnamefont{T.}~\bibnamefont{Jennewein}},
  \bibinfo{author}{\bibfnamefont{J.-W.} \bibnamefont{Pan}},
  \bibinfo{author}{\bibfnamefont{K.}~\bibnamefont{Resch}}, \bibnamefont{and}
  \bibinfo{author}{\bibfnamefont{A.}~\bibnamefont{Zeilinger}},
  \bibinfo{journal}{Phys. Rev. Lett.} \textbf{\bibinfo{volume}{92}},
  \bibinfo{pages}{017902} (\bibinfo{year}{2004}).

\bibitem[{\citenamefont{Martini et~al.}(1996)\citenamefont{Martini, Giuseppe,
  and Marrocco}}]{demartini1996}
\bibinfo{author}{\bibfnamefont{F.~D.} \bibnamefont{Martini}},
  \bibinfo{author}{\bibfnamefont{G.~D.} \bibnamefont{Giuseppe}},
  \bibnamefont{and} \bibinfo{author}{\bibfnamefont{M.}~\bibnamefont{Marrocco}},
  \bibinfo{journal}{Phys. Rev. Lett.} \textbf{\bibinfo{volume}{76}},
  \bibinfo{pages}{900} (\bibinfo{year}{1996}).

\bibitem[{\citenamefont{Brunel et~al.}(1999)\citenamefont{Brunel, Lounis,
  Tamarat, and Orrit}}]{brunel1999}
\bibinfo{author}{\bibfnamefont{C.}~\bibnamefont{Brunel}},
  \bibinfo{author}{\bibfnamefont{B.}~\bibnamefont{Lounis}},
  \bibinfo{author}{\bibfnamefont{P.}~\bibnamefont{Tamarat}}, \bibnamefont{and}
  \bibinfo{author}{\bibfnamefont{M.}~\bibnamefont{Orrit}},
  \bibinfo{journal}{Phys. Rev. Lett.} \textbf{\bibinfo{volume}{83}},
  \bibinfo{pages}{2722} (\bibinfo{year}{1999}).

\bibitem[{\citenamefont{Lounis and Moerner}(2000)}]{lounis2000}
\bibinfo{author}{\bibfnamefont{B.}~\bibnamefont{Lounis}} \bibnamefont{and}
  \bibinfo{author}{\bibfnamefont{W.~E.} \bibnamefont{Moerner}},
  \bibinfo{journal}{Nature (London)} \textbf{\bibinfo{volume}{407}},
  \bibinfo{pages}{491} (\bibinfo{year}{2000}).

\bibitem[{\citenamefont{Michler et~al.}(2000)\citenamefont{Michler, Kiraz,
  Becher, Schoenfeld, Petroff, Zhang, Hu, and Imamo\=glu}}]{michler2000}
\bibinfo{author}{\bibfnamefont{P.}~\bibnamefont{Michler}},
  \bibinfo{author}{\bibfnamefont{A.}~\bibnamefont{Kiraz}},
  \bibinfo{author}{\bibfnamefont{C.}~\bibnamefont{Becher}},
  \bibinfo{author}{\bibfnamefont{W.~V.} \bibnamefont{Schoenfeld}},
  \bibinfo{author}{\bibfnamefont{P.~M.} \bibnamefont{Petroff}},
  \bibinfo{author}{\bibfnamefont{L.}~\bibnamefont{Zhang}},
  \bibinfo{author}{\bibfnamefont{E.}~\bibnamefont{Hu}}, \bibnamefont{and}
  \bibinfo{author}{\bibfnamefont{A.}~\bibnamefont{Imamo\=glu}},
  \bibinfo{journal}{Science} \textbf{\bibinfo{volume}{290}},
  \bibinfo{pages}{2282} (\bibinfo{year}{2000}).

\bibitem[{\citenamefont{Santori et~al.}(2001)\citenamefont{Santori, Pelton,
  Solomon, Dale, and Yamamoto}}]{santori2001}
\bibinfo{author}{\bibfnamefont{C.}~\bibnamefont{Santori}},
  \bibinfo{author}{\bibfnamefont{M.}~\bibnamefont{Pelton}},
  \bibinfo{author}{\bibfnamefont{G.}~\bibnamefont{Solomon}},
  \bibinfo{author}{\bibfnamefont{Y.}~\bibnamefont{Dale}}, \bibnamefont{and}
  \bibinfo{author}{\bibfnamefont{Y.}~\bibnamefont{Yamamoto}},
  \bibinfo{journal}{Phys. Rev. Lett.} \textbf{\bibinfo{volume}{86}},
  \bibinfo{pages}{1502} (\bibinfo{year}{2001}).

\bibitem[{\citenamefont{Beveratos et~al.}(2002)\citenamefont{Beveratos,
  K\"{u}hn, Brouri, Gacoin, Poizat, and Grangier}}]{beveratos2002a}
\bibinfo{author}{\bibfnamefont{A.}~\bibnamefont{Beveratos}},
  \bibinfo{author}{\bibfnamefont{S.}~\bibnamefont{K\"{u}hn}},
  \bibinfo{author}{\bibfnamefont{R.}~\bibnamefont{Brouri}},
  \bibinfo{author}{\bibfnamefont{T.}~\bibnamefont{Gacoin}},
  \bibinfo{author}{\bibfnamefont{J.-P.} \bibnamefont{Poizat}},
  \bibnamefont{and} \bibinfo{author}{\bibfnamefont{P.}~\bibnamefont{Grangier}},
  \bibinfo{journal}{Eur. Phys. J. D} \textbf{\bibinfo{volume}{18}},
  \bibinfo{pages}{191} (\bibinfo{year}{2002}).

\bibitem[{\citenamefont{Kiraz et~al.}(2004)\citenamefont{Kiraz, Atat\"ure, and
  Imamo\=glu}}]{kiraz2004}
\bibinfo{author}{\bibfnamefont{A.}~\bibnamefont{Kiraz}},
  \bibinfo{author}{\bibfnamefont{M.}~\bibnamefont{Atat\"ure}},
  \bibnamefont{and}
  \bibinfo{author}{\bibfnamefont{A.}~\bibnamefont{Imamo\=glu}},
  \bibinfo{journal}{Phys. Rev. A} \textbf{\bibinfo{volume}{69}},
  \bibinfo{pages}{032305} (\bibinfo{year}{2004}).

\bibitem[{\citenamefont{Bayer and Forchel}(2002)}]{bayer2002}
\bibinfo{author}{\bibfnamefont{M.}~\bibnamefont{Bayer}} \bibnamefont{and}
  \bibinfo{author}{\bibfnamefont{A.}~\bibnamefont{Forchel}},
  \bibinfo{journal}{Phys. Rev. B} \textbf{\bibinfo{volume}{65}},
  \bibinfo{pages}{041308} (\bibinfo{year}{2002}).

\bibitem[{\citenamefont{Kammerer et~al.}(2002)\citenamefont{Kammerer,
  Cassabois, Perrin, Delalande, Roussignol, and G\'{e}rard}}]{kammerer2002}
\bibinfo{author}{\bibfnamefont{C.}~\bibnamefont{Kammerer}},
  \bibinfo{author}{\bibfnamefont{G.}~\bibnamefont{Cassabois}},
  \bibinfo{author}{\bibfnamefont{M.}~\bibnamefont{Perrin}},
  \bibinfo{author}{\bibfnamefont{C.}~\bibnamefont{Delalande}},
  \bibinfo{author}{\bibfnamefont{P.}~\bibnamefont{Roussignol}},
  \bibnamefont{and} \bibinfo{author}{\bibfnamefont{J.~M.}
  \bibnamefont{G\'{e}rard}}, \bibinfo{journal}{Appl. Phys. Lett.}
  \textbf{\bibinfo{volume}{81}}, \bibinfo{pages}{2737} (\bibinfo{year}{2002}).

\bibitem[{\citenamefont{Santori et~al.}(2002)\citenamefont{Santori, Fattal,
  Vu\u{c}kovi\'{c}, Solomon, and Yamamoto}}]{santori2002}
\bibinfo{author}{\bibfnamefont{C.}~\bibnamefont{Santori}},
  \bibinfo{author}{\bibfnamefont{D.}~\bibnamefont{Fattal}},
  \bibinfo{author}{\bibfnamefont{J.}~\bibnamefont{Vu\u{c}kovi\'{c}}},
  \bibinfo{author}{\bibfnamefont{G.~S.} \bibnamefont{Solomon}},
  \bibnamefont{and} \bibinfo{author}{\bibfnamefont{Y.}~\bibnamefont{Yamamoto}},
  \bibinfo{journal}{Nature (London)} \textbf{\bibinfo{volume}{419}},
  \bibinfo{pages}{594} (\bibinfo{year}{2002}).

\bibitem[{\citenamefont{Fattal et~al.}(2004{\natexlab{a}})\citenamefont{Fattal,
  Inoue, Vu\u{c}kovi\'{c}, Santori, Solomon, and Yamamoto}}]{fattal2004a}
\bibinfo{author}{\bibfnamefont{D.}~\bibnamefont{Fattal}},
  \bibinfo{author}{\bibfnamefont{K.}~\bibnamefont{Inoue}},
  \bibinfo{author}{\bibfnamefont{J.}~\bibnamefont{Vu\u{c}kovi\'{c}}},
  \bibinfo{author}{\bibfnamefont{C.}~\bibnamefont{Santori}},
  \bibinfo{author}{\bibfnamefont{G.~S.} \bibnamefont{Solomon}},
  \bibnamefont{and} \bibinfo{author}{\bibfnamefont{Y.}~\bibnamefont{Yamamoto}},
  \bibinfo{journal}{Phys. Rev. Lett.} \textbf{\bibinfo{volume}{92}},
  \bibinfo{pages}{037903} (\bibinfo{year}{2004}{\natexlab{a}}).

\bibitem[{\citenamefont{Fattal et~al.}(2004{\natexlab{b}})\citenamefont{Fattal,
  Diamanti, Inoue, and Yamamoto}}]{fattal2004b}
\bibinfo{author}{\bibfnamefont{D.}~\bibnamefont{Fattal}},
  \bibinfo{author}{\bibfnamefont{E.}~\bibnamefont{Diamanti}},
  \bibinfo{author}{\bibfnamefont{K.}~\bibnamefont{Inoue}}, \bibnamefont{and}
  \bibinfo{author}{\bibfnamefont{Y.}~\bibnamefont{Yamamoto}},
  \bibinfo{journal}{Phys. Rev. Lett.} \textbf{\bibinfo{volume}{92}},
  \bibinfo{pages}{037904} (\bibinfo{year}{2004}{\natexlab{b}}).

\bibitem[{\citenamefont{Jelezko et~al.}(2003)\citenamefont{Jelezko, Volkmer,
  Popa, Rebane, and Wratchup}}]{jelezko2003}
\bibinfo{author}{\bibfnamefont{F.}~\bibnamefont{Jelezko}},
  \bibinfo{author}{\bibfnamefont{A.}~\bibnamefont{Volkmer}},
  \bibinfo{author}{\bibfnamefont{I.}~\bibnamefont{Popa}},
  \bibinfo{author}{\bibfnamefont{K.~K.} \bibnamefont{Rebane}},
  \bibnamefont{and} \bibinfo{author}{\bibfnamefont{J.}~\bibnamefont{Wratchup}},
  \bibinfo{journal}{Phys. Rev. A} \textbf{\bibinfo{volume}{67}},
  \bibinfo{pages}{041802} (\bibinfo{year}{2003}).

\bibitem[{\citenamefont{Tamarat et~al.}(2000)\citenamefont{Tamarat, Maali,
  Lounis, and Orrit}}]{tamarat2000}
\bibinfo{author}{\bibfnamefont{P.}~\bibnamefont{Tamarat}},
  \bibinfo{author}{\bibfnamefont{A.}~\bibnamefont{Maali}},
  \bibinfo{author}{\bibfnamefont{B.}~\bibnamefont{Lounis}}, \bibnamefont{and}
  \bibinfo{author}{\bibfnamefont{M.}~\bibnamefont{Orrit}}, \bibinfo{journal}{J.
  Phys. Chem. A} \textbf{\bibinfo{volume}{104}}, \bibinfo{pages}{1}
  (\bibinfo{year}{2000}).

\bibitem[{\citenamefont{Moerner}(2002)}]{moerner2002}
\bibinfo{author}{\bibfnamefont{W.~E.} \bibnamefont{Moerner}},
  \bibinfo{journal}{J. Phys. Chem. B} \textbf{\bibinfo{volume}{106}},
  \bibinfo{pages}{910} (\bibinfo{year}{2002}).

\bibitem[{\citenamefont{Nonn and Plakhotnik}(2001)}]{nonn2001}
\bibinfo{author}{\bibfnamefont{T.}~\bibnamefont{Nonn}} \bibnamefont{and}
  \bibinfo{author}{\bibfnamefont{T.}~\bibnamefont{Plakhotnik}},
  \bibinfo{journal}{Chem. Phys. Lett.} \textbf{\bibinfo{volume}{336}},
  \bibinfo{pages}{97} (\bibinfo{year}{2001}).

\bibitem[{\citenamefont{Kiraz et~al.}(2003)\citenamefont{Kiraz, Ehrl,
  Br\"{a}uchle, and Zumbusch}}]{kiraz2003}
\bibinfo{author}{\bibfnamefont{A.}~\bibnamefont{Kiraz}},
  \bibinfo{author}{\bibfnamefont{M.}~\bibnamefont{Ehrl}},
  \bibinfo{author}{\bibfnamefont{C.}~\bibnamefont{Br\"{a}uchle}},
  \bibnamefont{and} \bibinfo{author}{\bibfnamefont{A.}~\bibnamefont{Zumbusch}},
  \bibinfo{journal}{J. Chem. Phys.} \textbf{\bibinfo{volume}{118}},
  \bibinfo{pages}{10821} (\bibinfo{year}{2003}).

\bibitem[{\citenamefont{Mais et~al.}(1999)\citenamefont{Mais, Basch\'{e},
  M\"uller, M\"ullen, and Br\"auchle}}]{mais1999}
\bibinfo{author}{\bibfnamefont{S.}~\bibnamefont{Mais}},
  \bibinfo{author}{\bibfnamefont{T.}~\bibnamefont{Basch\'{e}}},
  \bibinfo{author}{\bibfnamefont{G.}~\bibnamefont{M\"uller}},
  \bibinfo{author}{\bibfnamefont{K.}~\bibnamefont{M\"ullen}}, \bibnamefont{and}
  \bibinfo{author}{\bibfnamefont{C.}~\bibnamefont{Br\"auchle}},
  \bibinfo{journal}{Chem. Phys.} \textbf{\bibinfo{volume}{247}},
  \bibinfo{pages}{41} (\bibinfo{year}{1999}).

\end{thebibliography}

\newpage
\begin{figure}
\begin{center}
  \centerline{\psfig{figure=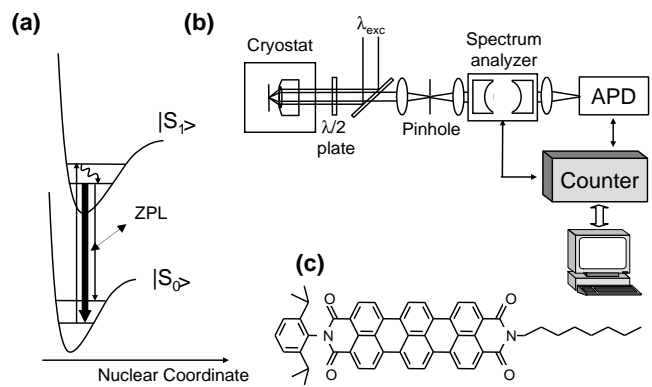,width=8.5cm,angle=0}}
  \caption{(a) Vibronic excitation scheme. (b) Experimental setup showing the Fabry-Perot spectrum analyzer used for linewidth measurements.
  (c) Terrylenediimide (TDI).}
\label{fig1}
\end{center}
\end{figure}

\newpage
\begin{figure}
\begin{center}
  \centerline{\psfig{figure=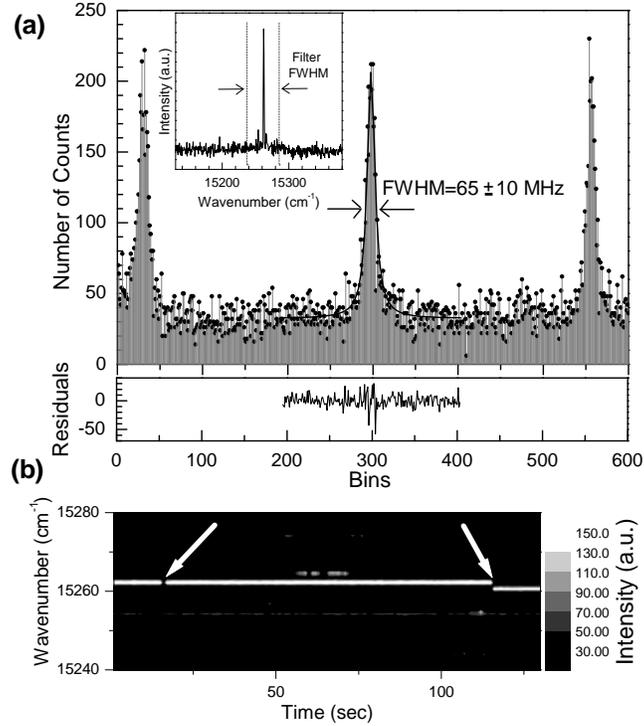,width=8.5cm,angle=0}}
  \caption{Emission spectra of the purely electronic zero-phonon line of a single TDI molecule in hexadecane excited at 605~nm and recorded with different resolutions. (a) Scan of the spectrum analyzer with 500~ms integration time per bin (separated by 2~ms dark periods). Excitation intensity of 35~$\mu$W. Inset: Dispersed fluorescence spectrum
after the interference filter. Excitation power of 1~mW, close to saturation. (b) Spectral trace with 10~mW excitation power. Blinking at t=16~sec and a spectral jump at t=115~sec are visible.}
\label{fig2}
\end{center}
\end{figure}

\newpage
\begin{figure}
\begin{center}
  \centerline{\psfig{figure=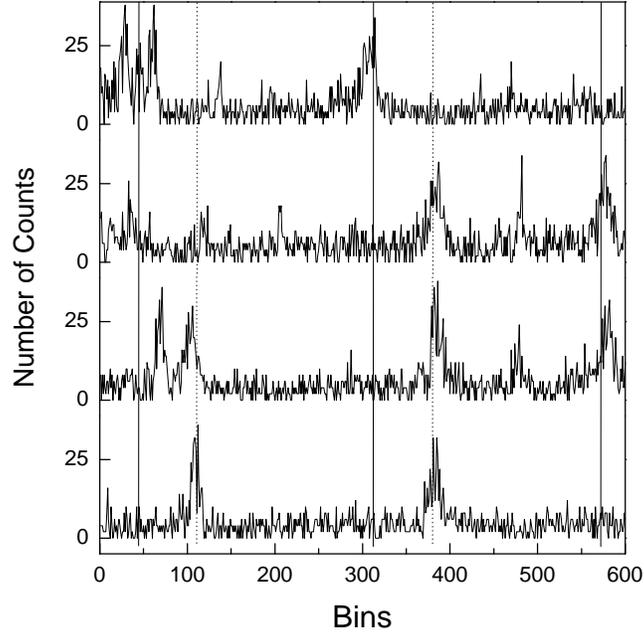,width=8.5cm,angle=0}}
  \caption{Consecutive scans of the spectrum analyzer for a single molecule's purely electronic zero-phonon line emission with a 25~ms integration time per bin (separated by 2~ms dark periods). Excitation at 608~nm with an intensity of 200~$\mu$W. Straight and dotted lines are repeated at 1.5~GHz intervals.}
\label{fig3}
\end{center}
\end{figure}

\end{document}